\documentclass[letterpaper]{article} 
\usepackage{aaai24}  
\nocopyright 
\usepackage{times}  
\usepackage{helvet}  
\usepackage{courier}  
\usepackage[hyphens]{url}  
\usepackage{graphicx} 
\usepackage{natbib}  
\usepackage{caption} 
\usepackage{algorithm}
\usepackage{algorithmic}
\usepackage{multirow}
\usepackage{subcaption}

\usepackage{amssymb}
\usepackage{amsthm, amsmath}
\theoremstyle{definition}

\usepackage{newfloat}
\usepackage{listings}
\DeclareCaptionStyle{ruled}{labelfont=normalfont,labelsep=colon,strut=off} 
\floatstyle{ruled}
\newfloat{listing}{tb}{lst}{}
\floatname{listing}{Listing}
\title{Workforce pDEI: Productivity Coupled with DEI}
\author{
    Lanqing Du\thanks{ld695@drexel.edu}, Jinwook Lee
}
\affiliations{
    Decision Sciences and MIS, Drexel University\\

    3220 Market St, Philadelphia, PA 19014
}

\usepackage{bibentry}

\begin{document}

\maketitle

\begin{abstract}
Ranking pertaining to the human-centered tasks -- underscoring their paramount significance in these domains such as evaluation and hiring process -- exhibits widespread prevalence across various industries. Consequently, decision-makers are taking proactive measurements to promote diversity, underscore equity, and advance inclusion. Their unwavering commitment to these ideals emanates from the following convictions: (i) Diversity encompasses a broad spectrum of differences; (ii) Equity involves the assurance of equitable opportunities; and (iii) Inclusion revolves around the cultivation of a sense of value and impartiality, concurrently empowering individuals. Data-driven AI tools have been used for screening and ranking processes. However, there is a growing concern that the presence of pre-existing biases in databases may be exacerbated, particularly in the context of imbalanced datasets or the black-box-schema. In this research, we propose a model-driven recruitment decision support tool that addresses fairness together with equity in the screening phase. We introduce the term ``pDEI" to represent the output-input oriented production efficiency adjusted by socioeconomic disparity. Taking into account various aspects of interpreting socioeconomic disparity, our goals are (i) maximizing the relative efficiency of underrepresented groups and (ii) understanding how socioeconomic disparity affects the cultivation of a DEI-positive workplace.

\end{abstract}

\section{Introduction}

In recent years, there has been a noticeable increase in discussions advocating for greater diversity, such as: ``Finance Functions Need More Women Leaders" \cite{news1}” and ``Why We Need More Faculty of Color in Higher Education \cite{news2}". These discussions have brought attention to underrepresented groups in the workforce, sparking further conversations about the root causes of the issue, the best timing for action, and effective approaches for addressing it.


The compelling necessity for addressing diversity issues lies in the potential organizational benefits that can arise from fostering DEI within the workforce. For example, diversified professional interactions can better serve a wider group of customers and increase sales and satisfaction rates \cite{Fenty, UCLA}. Integrating DEI into the workforce can enhance crisis communications, with a recommendation for the leadership team to exhibit cultural competence in communicating with diverse groups effectively \cite{hbrDEI}. 


As hiring with DEI becoming more crucial than ever, there is a pressing need for operational guidance on systematic recruitment to define and prioritize competitive applicants from underrepresented groups. Incorporating DEI into the hiring process may begin with the screening process, as it serves as the initial opportunity to address diversity, equity, and inclusion within the workforce.

\section{Problem Description}

Demographic factors like ethnicity, gender, and age can significantly impact personal experiences and professional interactions across various aspects of life. It's important to note that the ``entrance barrier" related to underrepresented groups can be context-dependent \cite{harris2015discrimination, fouad2017scct, AsianMovie}, varying not only between group stereotypes and individual stereotypes but also across different industries where the definition of ``underrepresented groups" may differ. The most ``overrepresented group" can often dominate the representation rate in various industries; nevertheless, there may be instances where multiple groups can be labeled as ``overrepresented" in certain niche industries while being considered ``underrepresented" in more mainstream sectors.

Consider a scenario where an organization is screening applicants based on their screening evaluations, rated on a scale of 1-7, with 7 being the highest. Suppose that there are two applicants, A (evaluation score 7) and B (evaluation score 6), competing for a single position. Lacking further context, applicant A may be favored as the superior candidate based on its higher evaluation score. Nevertheless, we should consider the scenario where applicant A belongs to an \textit{Overrepresented Group} while applicant B is from an \textit{Underrepresented Group} within the specific industry mentioned. At that point, hiring a more suitable candidate can be complicated, especially when considering related DEI values within the job description requirements.

\begin{figure}
    \centering
    \hspace*{-0.5cm}
    \includegraphics[width = 0.7\columnwidth, trim= 4cm 2cm 5cm 0cm, clip]{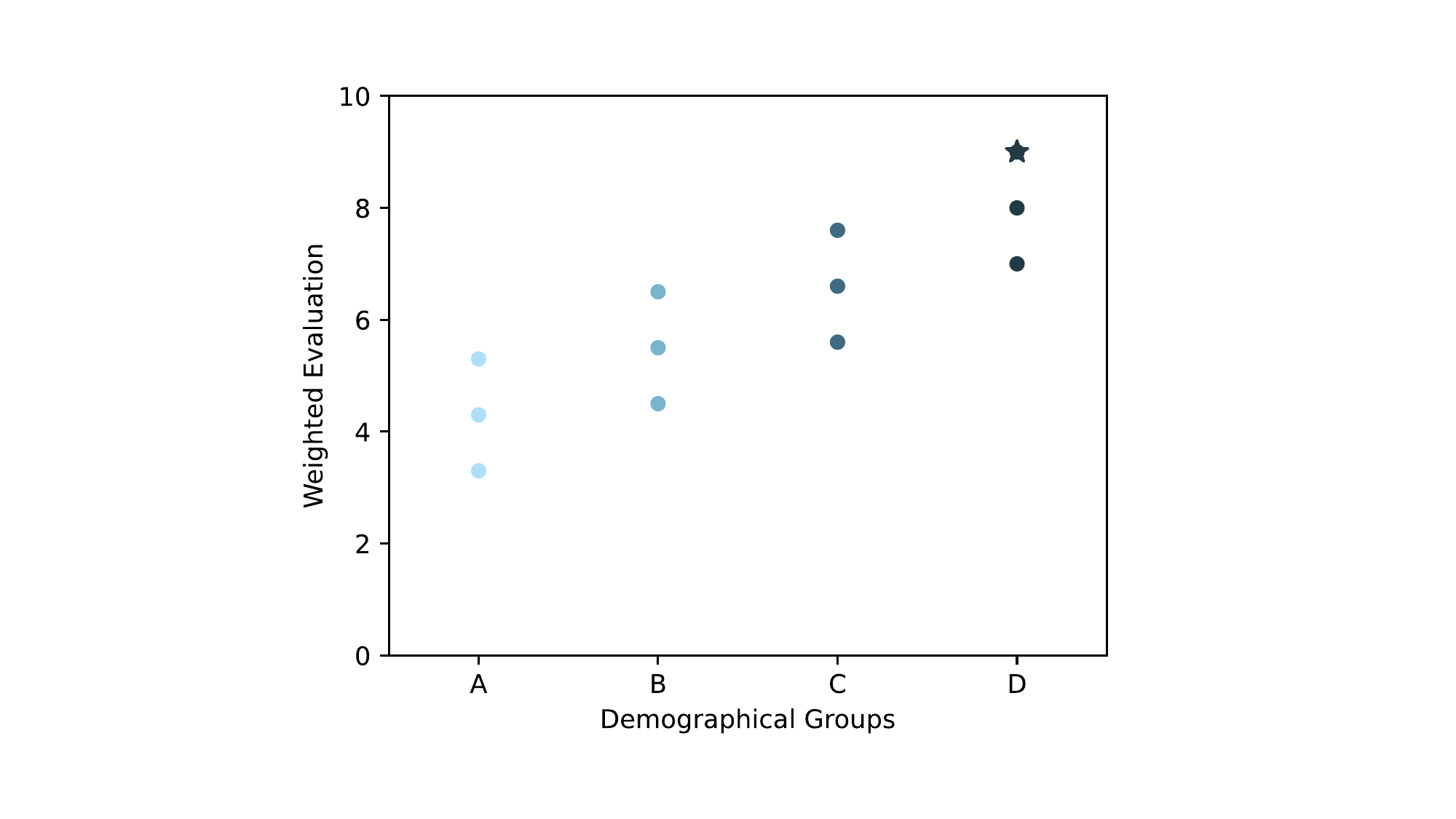}
    \caption{Evaluation by Demographic Groups.}
    \label{fig:noMetric}

\end{figure}

Assuming that Figure \ref{fig:noMetric} displays the top 3 candidates from each demographic group with their performance plotted on the same axis, it is evident that the individual marked with the star should be considered the top candidate across all applicant groups if without incorporating DEI consideration.

\begin{figure}
    \centering
      \begin{subfigure}{0.4\columnwidth}
        \centering
        {\hspace*{-1cm}\includegraphics[width = 1.7\columnwidth, trim=6cm 2cm 0cm 0cm, clip]{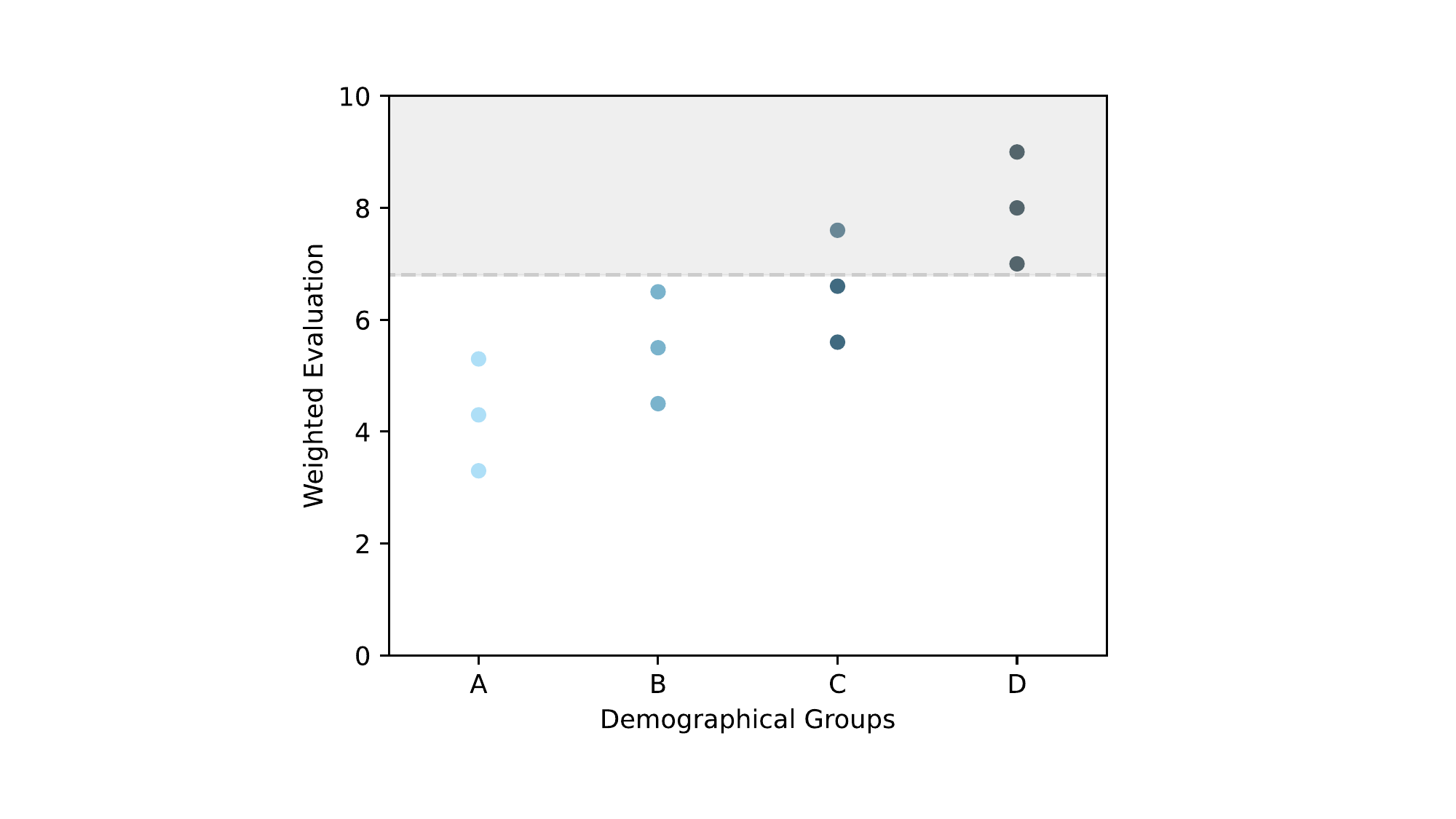}
        \caption{Selection scheme 1}}
        \label{fig:1-1}
      \end{subfigure}
      \begin{subfigure}{0.4\columnwidth}
        \centering
        {\hspace*{0cm}\includegraphics[width = 1.7\columnwidth, trim=6cm 2cm 0cm 0cm, clip]{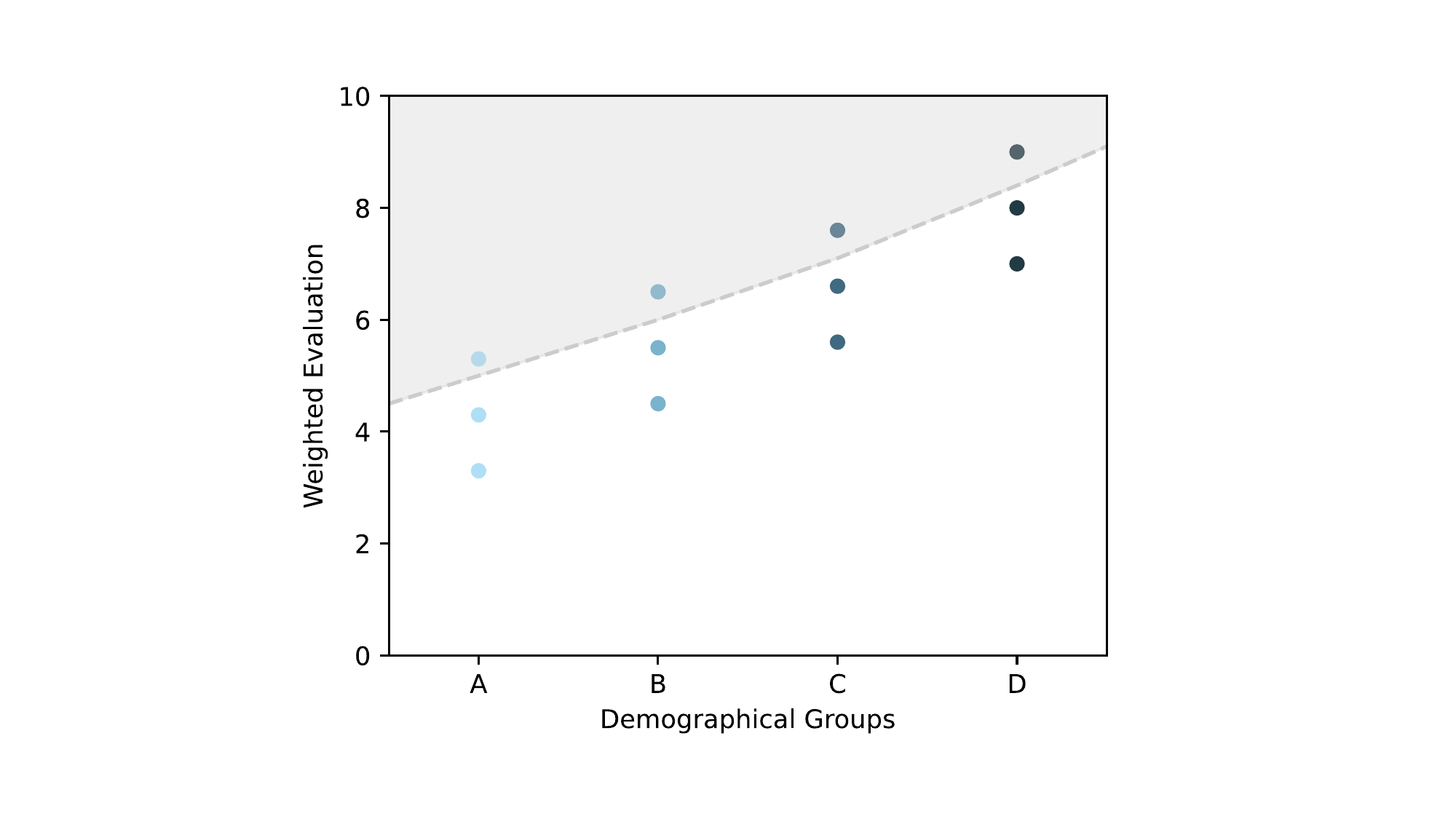}}
        \caption{Selection scheme 2}
        \label{fig:1-2}
      \end{subfigure} 
     
    \caption{Two Selection Schemes.}
    \label{fig:probDesPar}
\end{figure}

Assume that the job screening process aims to choose 4 candidates for the next round interview. Without taking DEI into account, the conventional approach would involve selecting individuals from the highlighted region characterized by higher evaluation scores in Figure \ref{fig:probDesPar}(a). Nevertheless, this selection scheme could potentially conflict with the goals of DEI, especially when groups A and B represent the most socioeconomically disadvantaged segments of the population. This conflict manifests in two ways: (1) the potential violation of the ``$4/5$ rule" as outlined in federal guidelines \cite{federal}, and (2) a concern that, in the long term, equity may prove insufficient in addressing underlying socioeconomic inequalities. 

One might perceive an approach akin to equity, as illustrated in Figure \ref{fig:probDesPar}(b), as a potential solution to control outcome disparities \cite{raghavan2020mitigating}. Nonetheless, it is imperative to thoroughly assess the extent to which this selection process can be justified and whether it would be regarded as equitable for applicants with higher evaluation but are not selected from other groups.

In addressing this question, it becomes paramount to provide a comprehensive justification and elucidation of the notions of \textit{Overrepresented} and \textit{Underrepresented} with regard to the factors influencing socioeconomic parity levels.

\begin{figure}
    \centering
      \begin{subfigure}{0.4\columnwidth}
        \centering
        {\hspace*{-1cm}\includegraphics[width = 1.9\columnwidth, trim=6cm 2cm 0cm 0cm, clip]{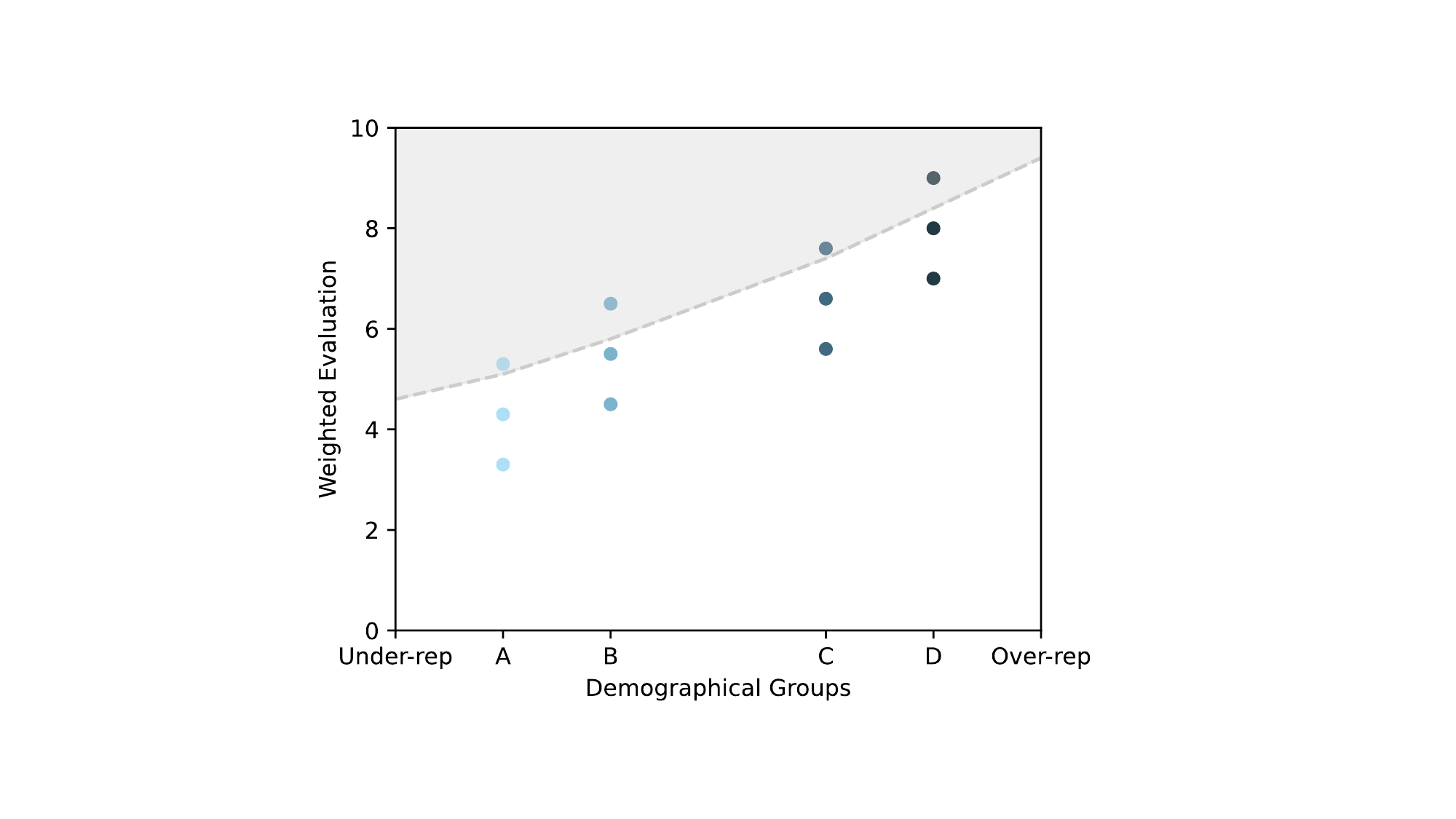}
        \caption{Scheme 2 case 1}}
        \label{fig:2-1}
      \end{subfigure}
      \begin{subfigure}{0.4\columnwidth}
        \centering
        {\hspace*{0cm}\includegraphics[width = 1.9\columnwidth, trim=6cm 2cm 0cm 0cm, clip]{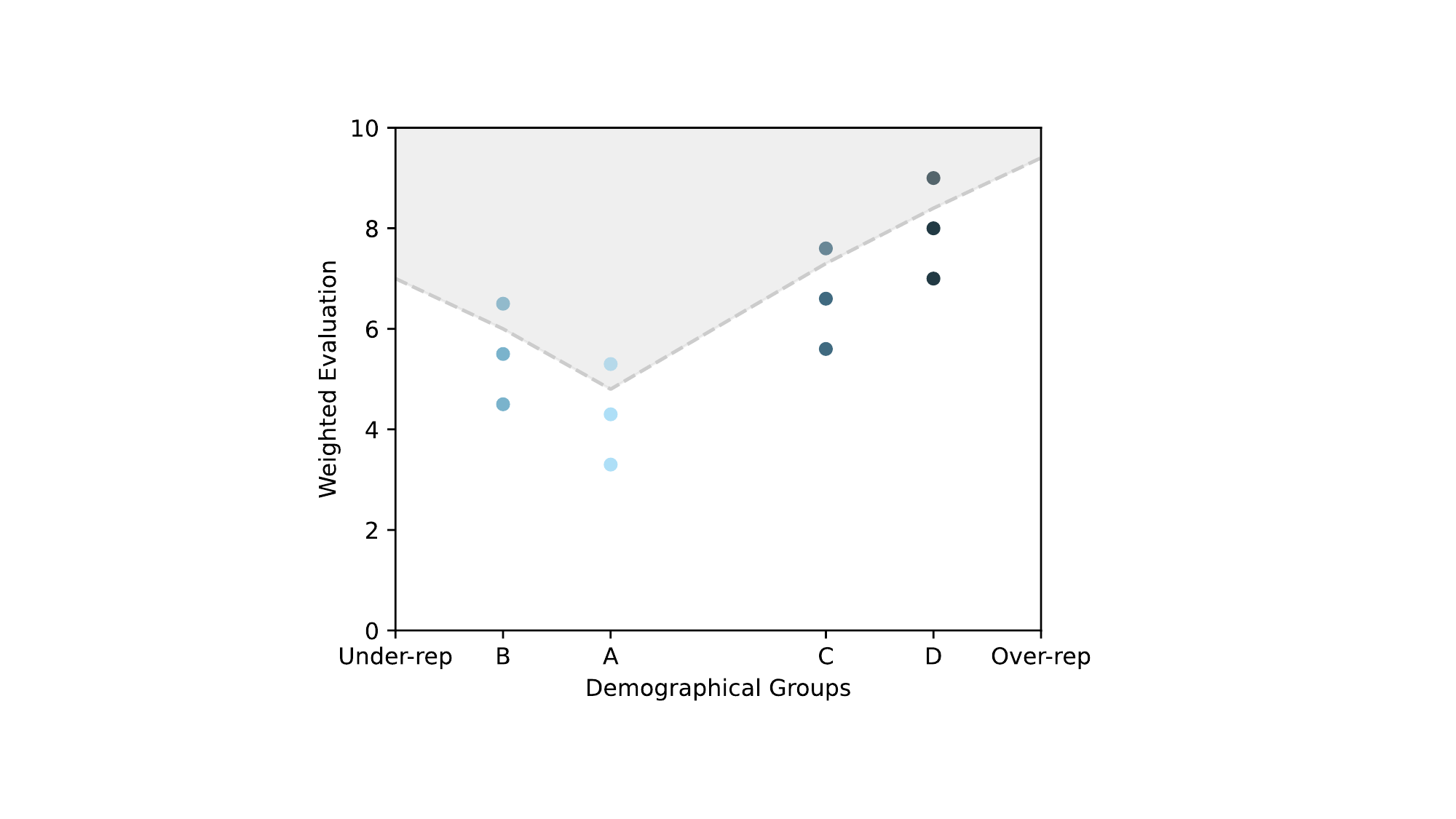}}
        \caption{Scheme 2 case 2}
        \label{fig:2-2}
      \end{subfigure} 
    \caption{Evaluation coupled with Socioeconomic Parity.}
    \label{fig:probDesPa$R_2$}
\end{figure}

As an illustration, consider the scenario where we have two distinct sets of ground truth labels representing socioeconomic disparity for each group, as depicted in Figure \ref{fig:probDesPa$R_2$}. These labels are arranged along a spectrum, representing a continuum from unfairness to fairness, with the interpretation being from (the most) underrepresented to (the most) overrepresented. If we apply the identical selection scheme (as illustrated in Figure \ref{fig:probDesPar}(b)), wherein the highlighted region signifies the same set of chosen candidates. The consistency of this selection approach appears more reasonable when considering the disparity as represented in case 1 (Figure \ref{fig:probDesPa$R_2$}(a)) rather than in case 2 (Figure \ref{fig:probDesPa$R_2$}(b)). As in case 2, the candidate chosen from group A may be perceived as having a comparative advantage over group B. However, it's noteworthy that the selected candidate from group A exhibits even lower evaluation scores than one candidate from group B who was not selected, making the selected group A candidate less competitive than the group B's second-highest candidate (unselected).

\subsection{Expected Outcomes and Benefits}

In order to solve the above-mentioned problems, we propose DEI-embedded recruitment  model that provide an optimal balance to promote DEI while ensuring overall competency. In which, we consider the output-input oriented efficiency (productivity) captured in the fraction of 
\begin{equation}\label{pDEI}
    pDEI = \frac{\text{Evaluation}}{\text{Socioeconomic Fairness}}
\end{equation}

\noindent to adjust candidates competency. With the proposed model, one can make a more informed decision-making with given requirements of competency and DEI.

Our numerical test demonstrates that the perceived conventional practice of selecting an equal number (or proportion) of applicants might not be deemed sufficiently equitable, which, in turn, depends on the interplay between parity measurements and evaluation scores. 

In a more comprehensive perspective, such knowledge can be transferred  to the different domain for AI-driven decision-making. This involves emphasizing the significance of understanding the underlying invariant sensitive features, which can offer valuable guidance when applying the model to other domain. It is also crucial to recognize that these models should not solely focus on addressing imbalanced dataset. In addition to considering the ``focal loss", the criteria for evaluating the overall model performance across various groups should be reweighted once more to account for a closer examination of socioeconomic disparities.


\section{Methodology and Literature Review}

\subsection{Bias and Fairness}

In the context of human-centered decision-making, bias is often associated with human bias \cite{bias} and can potentially manifest as algorithmic bias in AI tools \citep{dwork2018group, kordzadeh2022algorithmic}. 

The goal of mitigating bias arises from the critical need to ensure that models do not manifest discrimination against groups based on their protected attributes, such as demographic characteristics. However, when considering the DEI requirements, we do enforce that our model compensates the underrepresented groups, thus we focus on fairness under equitable considerations.  

In this context, we introduce fairness as a gauge of demographic groups' relative socioeconomic status. Utilizing a fairness metric, we discern individuals belonging to groups characterized by lower fairness scores, consequently highlighting pronounced disparities in socioeconomic inequality. Thus, within the context of Equation (\ref{pDEI}), for a given arbitrary constant in the numerator, a reduction in the denominator reflecting the disparity would equitably enhance the value of pDEI, taking into account the presence of socioeconomic inequalities.

\subsection{Parity and Fairness Metrics}

Without considering TPR and FPR, the derivations of the parity measurements briefly cover two concepts: \textit{Statistical Parity} \citep{zemel2013learning, dwork2018fairness} and \textit{Disparate Impact} \cite{barocas2016big}, both of which emphasize equal rates of favorable outcomes across overrepresented and underrepresented groups. When introducing TPR and FPR, the goal shifts to emphasize equality across overrepresented and underrepresented groups in the following ways: (1) Equal true positive rates in \textit{TPR Parity}, (2) Equal false positive rates in \textit{FPR Parity}, and (3) Equal true positive and false positive rates in \textit{Average Absolute Odds}.

An extensive list of fairness metrics can be found in \citep{EquityHooker,chen2022combining}, which also explores mathematical properties and related optimizations of these metrics.

\begin{table}[htbp]
    \centering

    \begin{small}
    \begin{tabular}{p{3cm}p{4.7cm}}
    \hline 
     Metric & Formulas \\
    \hline 
    Statistical Parity & $\scriptstyle  p(\mathbf{a} = 1 | \ \mathbf{b} = b) = p(\mathbf{a} = 1 | \ \mathbf{b} = b') $\\
    Disparate Impact &  $\scriptstyle  p(\mathbf{a} = 1 | \ \mathbf{b} = b) / p(\mathbf{a} = 1 | \ \mathbf{b} = b') $ \\
    TPR Parity & $\scriptstyle  TPR_{\mathbf{b} = b}  = TPR_{\mathbf{b} = b'} $\\
    FPR Parity & $\scriptstyle  FPR_{\mathbf{b} = b} = FPR_{\mathbf{b} = b'} $ \\
    Average Absolute Odds &  $\scriptstyle  \frac{|TPR_{\mathbf{b} = b}-TPR_{\mathbf{b} = b'}| + |FPR_{\mathbf{b} = b}-FPR_{\mathbf{b} = b'}|}{2}$ \\
    \hline
    \end{tabular}
    \end{small}
    \caption{Fairness Metrics: Parity Measures}
    \label{tab:my_label}
\end{table}


In this research, we use Disparate Impact (DI, as referenced later) by considering the fraction of conditional probabilities due to its property of yielding nonnegative results:
\begin{equation}\label{DI}
    p(\mathbf{a} = 1 | \ \mathbf{b} = b) / p(\mathbf{a} = 1 | \ \mathbf{b} = b')
\end{equation}

\noindent where $\mathbf{a}$ represents an outcome variable and $\mathbf{b}$ denotes the examined group as defined by the protected features. 



\subsection{DEA Models}

Data envelopment analysis (DEA) models are widely employed for assessing a flexible ``fair" form of efficiency in situations involving multiple inputs and multiple outputs \cite{cook2009data, khodabakhshi2014fair, zhu2021cross, kremantzis2022fairer}. 

DEA models originated from the CRS model, in which \cite{charnes1978measuring} proposed the economic term of efficiency for a given decision-making unit with input $x_0$ and output $y_0$ relative to the industry-level benchmark $\tilde{y}$ to match the engineering efficiency of $y_0 / \tilde{y}$, where, given a set of decision making units (DMUs) $I$, treating the industry benchmark $\tilde{y}$ as one of the units from the given group $I$, then we can measure the relative efficiency for $\text{DMU}_0$ with input $x_0$ and output $y_0$:

\begin{equation}\label{eq1}
    \begin{array}{ll}    
        max_{\mu,\nu} & \cfrac{\mu y_0}{\nu x_0}   \\[1em]
        s.t.  &  \cfrac{\mu y_i}{ \nu x_i}\leq 1, \forall i \in I,\\[1em]
         &  \mu,\nu \geq 0. \\
    \end{array}
\end{equation}

Switching to the multiple inputs outputs scenario, let us denote a given DMU $i$ with the input vector: $\mathbf{x}_i = (x_{i1}, x_{i2}, \cdots, x_{ir})$, and output vector $\mathbf{y}_i = (y_{i1}, y_{i2}, \cdots, y_{is})$, Model (\ref{eq1}) could be further written into the following linear form:

\begin{equation}\label{eq2}
    \begin{array}{ll}
        max_{\mu,v} & \sum_{p=1}^{s} \mu_p y_{0p}\\ [1em]
        s.t.          & \sum_{q=1}^{r} \nu_q x_{0q} = 1 \\ [1em]
        &  \sum_{p=1}^{s} \mu_p y_{ip}-\sum_{q=1}^{r} \nu_qx_{iq}\leq 0, \forall i \in I, \\ [1em]

         &  \mu, \nu \geq 0. \\
    \end{array}
\end{equation}

\noindent where the pivotal idea of utilizing the weights assigned to a given DMU to gauge the efficiency of other DMUs holds significant promise for enhancing the efficiency of the specified unit. This interpretation aligns with the principles of DEI, as it aims to enhance the scores of the designated unit while assuming that other units can achieve at least the same level of efficiency using identical weights.

In the context of AI-driven screening processes, vendors typically furnish an evaluation that rates applicants across a spectrum of competencies, often represented as multi-dimensional assessment outcomes \cite{raghavan2020mitigating}. These multi-dimensional outcomes would be fed into a model for assessing the candidates' competency, making them appropriate for the application of DEA models.

\subsection{pDEI}

In Definition (\ref{pDEI}), it becomes imperative to refine the evaluation of candidates by considering their socioeconomic disparities. In the context of this study, we utilize labor statistics as essential information to analyze and assess the socioeconomic disparities related to job screening tasks, with the aim of elucidating the underlying assumptions connecting the ``entrance barrier" and ``socioeconomic disparities."

In light of the constrained dataset granularity within each subcategory, we adopt a method of cross-referencing statistics \#10 \cite{laborStat10} and \#11 \cite{laborStat11} to enrich our comprehension of disparity metrics for various demographic groups. Subsequently, we integrate this enriched knowledge into Model (\ref{eq2}) as an input vector, enhancing the model's capacity to account for variations in disparity across groups. By fixing the relative fairness scores generated from the labor statistics within each industry, we establish a fixed socioeconomic fairness metric $\mathbf{x}_i$ for each group based on their characteristics derived from labor statistics. Please note that the cardinality of $\mathbf{x}_i$ holds significant importance when determining which features to include in order to further address fairness considerations.

\section{Numerical Findings}

\subsection{Labor Statistics Fact}

Our input vector centers on the conditional probability $p(\mathbf{a} = 1 | \ \mathbf{b} = b)$, representing our assessment of the employment rate, i.e. the event of hired ($\mathbf{a} = 1$) given applicant from a certain demographic group ($\mathbf{b} = b$), which can further be referenced as parity within a specific demographic group. In this evaluation, the specified labor force criteria of ``Total, 16 years and over," as outlined in \cite{laborStat10}, have been employed to approximate the overall available workforce within a specific cohort. 

It is imperative to acknowledge that, due to constraints imposed by page limitations, we have incorporated six industry sectors, primarily focusing on those within the ``Management, business, and financial operations occupations" category. These sectors have been included based on two considerations: (1) the total employed population of over 500,000 individuals and (2) a combined proportion from each demographic group not exceeding 110\%, a criterion designed to account for the influence of individuals from more than two racial backgrounds without unduly overshadowing the impact of other racial groups.

Table \ref{DI} displays the conditional probabilities $p(\mathbf{b} = b | \ \mathbf{a} = 1 )$, representing the percentages of employees from specific group $b = R_i$ within each sector. The row labels denote distinct sectors within the ``Management, Business, and Financial Operations Occupations": $S_1$ represents Chief executives, $S_2$ represents Sales managers Computer and information systems managers, $S_3$ represents Medical and health services managers, $S_4$ represents Education and childcare administrators Property,  $S_5$ represents Real estate, and $S_6$ represents Community association managers. For the column labels, ``Total" denotes the employment figures for a given sector (in thousands); ``Percentage" signifies the employment rate for a given group within the entire workforce of that sector. Among the groups, ``$G_1$" represents women, ``$R_1$" represents White, ``$R_2$" represents Black or African American, ``$R_3$" represents Asian, and ``$R_4$" represents Hispanic or Latino ethnicity."  Since the labor statistics exclusively provide the employment percentage for the group of women, we designate the non-woman group as ``$G_2$" in Table (\ref{tab:DI}).

\begin{table}[htbp]
  \centering

  \begin{small}
    \begin{tabular}{c|c|c|c|c|cc}
    \hline
    \multirow{2}[4]{*}{Industry} & \multirow{2}[4]{*}{Total} & \multicolumn{5}{c}{Percentage} \\
    \cline{3-7}          &       & $G_1$    & \multicolumn{1}{c|}{$R_1$} & \multicolumn{1}{c|}{$R_2$} & $R_3$    & $R_4$ \\
    
    \hline
     $S_1$& 1780  & 29.2  & 85.9  & 5.9   & 6.7   & 6.8 \\
     $S_2$& 566   & 34.2  & 88.3  & 5.9   & 3.5   & 11.3 \\
     $S_3$& 764   & 26.4  & 72.6  & 7.8   & 16.5  & 7.5 \\
     $S_4$& 797   & 71.6  & 74.6  & 16.0  & 7.3   & 9.0 \\
     $S_5$& 988   & 68.1  & 78.0  & 16.3  & 4.0   & 9.8 \\
     $S_6$& 835   & 50.3  & 83.1  & 9.9   & 4.3   & 11 \\
    \hline
    \end{tabular}%
  \end{small}
  \caption{Employed Persons by Detailed Occupation \cite{laborStat11}.}
  \label{tab:stat1}%
\end{table}%

It is worth noting that a substantial portion of the statistics is reported in a percentage format. Consequently, it is important to acknowledge the potential for variations in the presented results attributable to the propagation of errors during calculations.

\begin{figure}[htbp]
    \centering
    {\hspace*{-0cm}\includegraphics[width = 1.3\columnwidth, trim=6cm 4cm 0cm 4cm, clip]{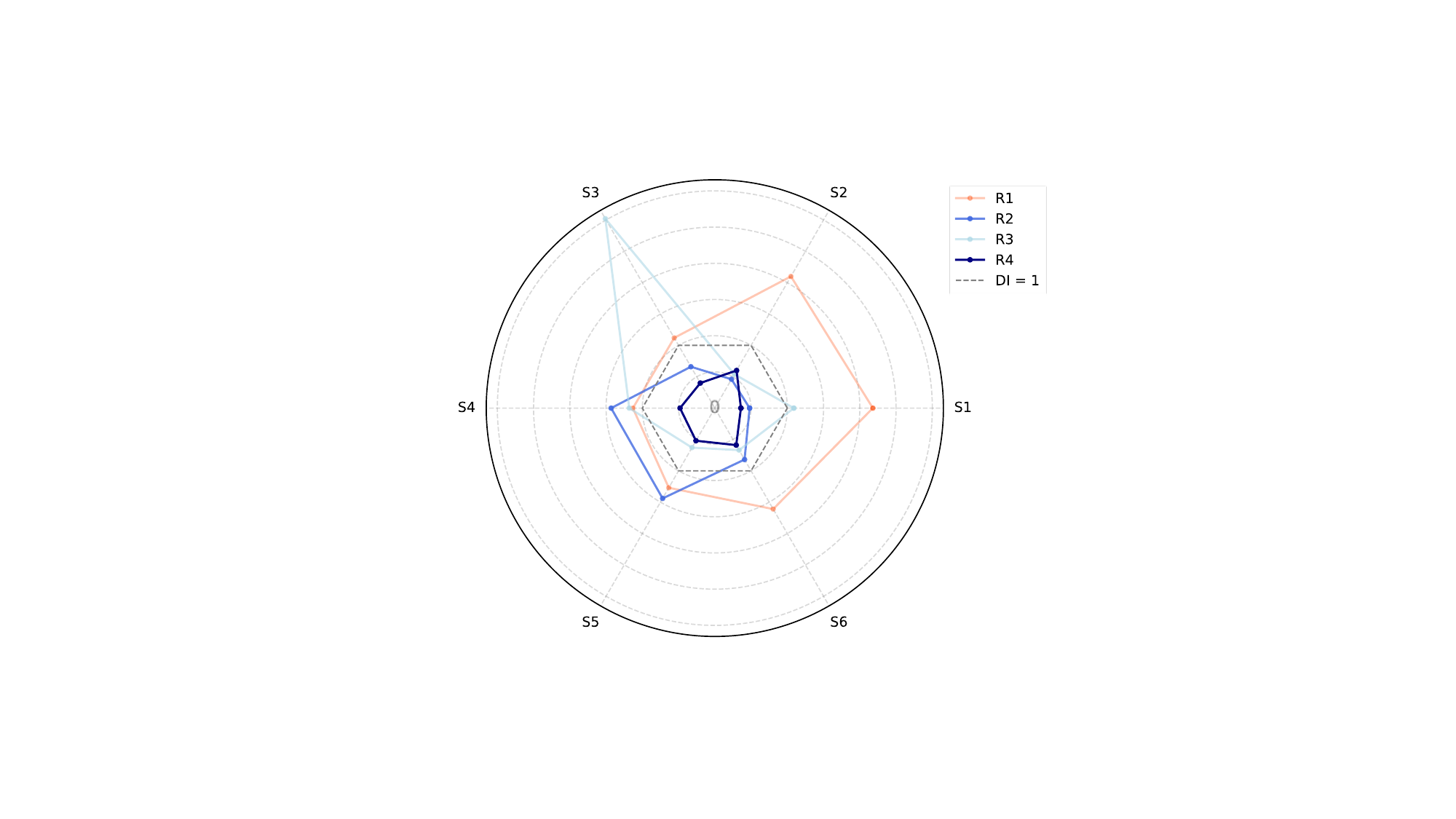}}
    \caption{Disparate Impact by Industry and Race/Ethnicity.}
    \label{fig:DI}
\end{figure}

By cross-referencing the total number of employed individuals within each group, as documented in \cite{laborStat10}—121,908 from group $R_1$; 19,937 from $R_2$; 10,615 from $R_3$; and 29,299 from $R_4$—we calculate the disparate impact for each group across the above-mentioned six sectors, as in Table \ref{tab:DI} and Figure \ref{fig:DI}:

\begin{table}[htbp]
    \centering
    \begin{small}
    \begin{tabular}{l|rrrr|rr}
    \hline
          & \multicolumn{4}{c|}{Race \& Ethnicity} & \multicolumn{2}{c}{Gender} \\
          & \multicolumn{1}{l}{$R_1$} & \multicolumn{1}{l}{$R_2$} & \multicolumn{1}{l}{$R_3$} & \multicolumn{1}{l|}{$R_4$} & \multicolumn{1}{l}{$G_1$} & \multicolumn{1}{l}{$G_2$} \\
    \hline
    $S_1$    & 2.18  & 0.48  & 1.09  & 0.36  & 0.41  & 2.42 \\
    $S_2$    & 2.10  & 0.46  & 0.54  & 0.60  & 0.52  & 1.92 \\
    $S_3$    & 1.12  & 0.66  & 3.02  & 0.40  & 0.36  & 2.78 \\
    $S_4$    & 1.13  & 1.43  & 1.18  & 0.48  & 2.53  & 0.40 \\
    $S_5$    & 1.27  & 1.44  & 0.63  & 0.52  & 2.14  & 0.47 \\
    $S_6$    & 1.61  & 0.82  & 0.67  & 0.59  & 1.01  & 0.99 \\
    \hline
    \end{tabular}%
    \end{small} 
    \caption{Disparate Impact.}
    \label{tab:DI}
\end{table}

In Figure \ref{fig:DI}, the central point of the star plot represents the scenario where ``DI = 0," and it is connected by a dashed line to the point where ``DI = 1" across various sectors. It is evident that among the selected six sectors, ``$R_1$" exhibits the highest level of overrepresentation, with its DI values exceeding 1 across all sectors; while ``$R_4$" demonstrates the most significant underrepresentation across these industries. Furthermore, ``$R_2$" and ``$R_3$" are found to be overrepresented in certain sectors but underrepresented in the larger ones.

\subsection{Numerical Results}
In this study, we utilize the Matlab Toolbox developed by \citet{balk2021evaluation} for the integration of classical DEA models.

We investigate multiple evaluation scenarios to analyze the level of adjustment across diverse geographical groups. Nonetheless, due to data granularity constraints, our ability to compute the DI is constrained. Given the absence of the most disaggregated data of (Race, Gender) groups, we could only include two input variables for each candidate: (1) DI based on its race and ethnicity ($b \in \{ R_1, R_2, R_3, R_4 \}$), and (2) DI based on its gender ($b \in \{ G_1, G_2\}$).

\subsection{Uniform Evaluation Across Groups}

Suppose we are assessing the performance of the top four candidates from each demographic group, denoted as $C_{j,{R_i}}$: the $j^{th}$ top candidate from group $R_i$. For simplicity, we assign uniform scores across all race groups as follows: $C_{1,{R_i}}$ = (8, 8, 8, 8), $C_{2,{R_i}}$ = (7, 7, 7, 7), $C_{3,{R_i}}$ = (6, 6, 6, 6), and $C_{4,{R_i}}$ = (5, 5, 5, 5) for the first, second, third, and fourth top candidates across all race groups, respectively.

Tables \ref{tab:s1_uni}, \ref{tab:s2_uni}, \ref{tab:s5_uni}, and \ref{tab:s6_uni} display the adjusted efficiency score $pDEI$ obtained by treating DI as input in Equation (\ref{eq2}) across various sectors. In Tables \ref{tab:s1_uni} and \ref{tab:s2_uni}, the sectors exhibit underrepresentation of the group of women (DI $<$ 1), while the sectors in Tables \ref{tab:s5_uni} and \ref{tab:s6_uni} demonstrate overrepresentation of the group of women (DI $>$ 1).

\begin{table}[htbp]
    \centering
    \begin{small}
         \begin{tabular}{rccccc}
    \hline
          &       & C1    & C2    & C3    & C4 \\
    \hline
          & R1    & 0.17  & 0.14  & 0.12  & 0.10 \\
    \multicolumn{1}{c}{Race and } & R2     & 0.75 & 0.65  & 0.56  & 0.47 \\
    \multicolumn{1}{c}{Ethnicity} & R3     & 0.33  & 0.29  & 0.25  & 0.21 \\
          & R4     & 1 & 0.88 &0.75 & 0.63 \\
    \hline
          & R1  \& G1 & 1     & 0.88  & 0.75  & 0.63 \\
          & R2 \& G1 & 1     & 0.88  & 0.75  & 0.63 \\
    \multicolumn{1}{c}{Race and } & R3  \& G1 & 1     & 0.88  & 0.75  & 0.63 \\
    \multicolumn{1}{c}{Ethnicity} & R4  \& G1 & 1     & 0.88  & 0.75  & 0.63 \\
\cline{2-6}    \multicolumn{1}{c}{and} & R1 \& G2 & 0.17  & 0.15  & 0.13  & 0.11 \\
    \multicolumn{1}{c}{Gender} & R2 \& G2 & 0.75 & 0.65  & 0.56  & 0.47 \\
          & R3 \& G2 & 0.33  & 0.29  & 0.25  & 0.21 \\
          & R4 \& G2 & 1 & 0.88 & 0.75 & 0.63 \\
    \hline
    \end{tabular}%
    \end{small}
   
    \caption{$pDEI$ Scores in Industry $S_1$. }
    \label{tab:s1_uni}
\end{table}

\begin{table}[htbp]
    \centering
    \begin{small}
        \begin{tabular}{rccccc}
    \hline
          &       & C1    & C2    & C3    & C4 \\
    \hline
          & R1    & 0.22  & 0.19  & 0.16  & 0.14 \\
    \multicolumn{1}{c}{Race and } & R2    & 1.00  & 0.88  & 0.75  & 0.63 \\
    \multicolumn{1}{c}{Ethnicity} & R3    & 0.85  & 0.74  & 0.64  & 0.53 \\
          & R4    & 0.76  & 0.67  & 0.57  & 0.48 \\
    \hline
          & R1 \& G1 & 1     & 0.88  & 0.75  & 0.63 \\
          & R2 \& G1 & 1     & 0.88  & 0.75  & 0.63 \\
    \multicolumn{1}{c}{Race and } & R3 \& G1 & 1     & 0.88  & 0.75  & 0.63 \\
    \multicolumn{1}{c}{Ethnicity} & R4 \& G1 & 1     & 0.88  & 0.75  & 0.63 \\
\cline{2-6}    \multicolumn{1}{c}{and} & R1 \& G2 & 0     & 0.24  & 0.20  & 0.17 \\
    \multicolumn{1}{c}{Gender} & R2 \& G2 & 1     & 0.88  & 0.75  & 0.63 \\
          & R3 \& G2 & 1     & 0.74  & 0.64  & 0.53 \\
          & R4 \& G2 & 0.76  & 0.67  & 0.57  & 0.48 \\
    \hline
    \end{tabular}%
    \end{small}
    \caption{$pDEI$ Scores in Industry $S_2$. }
    \label{tab:s2_uni}
\end{table}

\begin{table}[htbp]
    \centering
    \begin{small}
    \begin{tabular}{rccccc}
    \hline
          &       & C1    & C2    & C3    & C4 \\
    \hline
          & R1    & 0.41  & 0.36  & 0.31  & 0.26 \\
    \multicolumn{1}{c}{Race and } & R2    & 0.36  & 0.32  & 0.27  & 0.23 \\
    \multicolumn{1}{c}{Ethnicity} & R3    & 0.83  & 0.72  & 0.62  & 0.52 \\
          & R4    & 1.00  & 0.88  & 0.75  & 0.63 \\
    \hline
          & R1 \& G1 & 0.41 & 0.36  & 0.31  & 0.26 \\
          & R2 \& G1 & 0.36 & 0.32  & 0.27  & 0.23 \\
    \multicolumn{1}{c}{Race and } & R3 \& G1 & 0.83 & 0.72  & 0.62  & 0.52 \\
    \multicolumn{1}{c}{Ethnicity} & R4 \& G1 & 1     & 0.88  & 0.75  & 0.63 \\
\cline{2-6}    \multicolumn{1}{c}{and} & R1 \& G2 & 1     & 0.88  & 0.75  & 0.63 \\
    \multicolumn{1}{c}{Gender} & R2 \& G2 & 1     & 0.88  & 0.75  & 0.63 \\
          & R3 \& G2 & 1     & 0.88  & 0.75  & 0.63 \\
          & R4 \& G2 & 1     & 0.88  & 0.75  & 0.63 \\
    \hline
    \end{tabular}%
    \end{small}
    
    \caption{$pDEI$ Scores in Industry $S_5$. }
    \label{tab:s5_uni}
\end{table}

\begin{table}[htbp]
    \centering
    \begin{small}
        \begin{tabular}{rccccc}
    \hline
          &       & C1    & C2    & C3    & C4 \\
    \hline
          & R1    & 0.36  & 0.32  & 0.27  & 0.23 \\
    \multicolumn{1}{c}{Race and } & R2    & 0.72  & 0.63  & 0.54  & 0.45 \\
    \multicolumn{1}{c}{Ethnicity} & R3    & 0.88  & 0.77  & 0.66  & 0.55 \\
          & R4    & 1.00  & 0.88  & 0.75  & 0.63 \\
    \hline
          & R1 \& G1 & 0.98  & 0.85  & 0.73  & 0.61 \\
          & R2 \& G1 & 0.98  & 0.85  & 0.73  & 0.61 \\
    \multicolumn{1}{c}{Race and } & R3 \& G1 & 0.98  & 0.85  & 0.73  & 0.61 \\
    \multicolumn{1}{c}{Ethnicity} & R4 \& G1 & 1     & 0.88  & 0.75  & 0.63 \\
\cline{2-6}    \multicolumn{1}{c}{and} & R1 \& G2 & 1     & 0.88  & 0.75  & 0.63 \\
    \multicolumn{1}{c}{Gender} & R2 \& G2 & 1     & 0.88  & 0.75  & 0.63 \\
          & R3 \& G2 & 1     & 0.88  & 0.75  & 0.63 \\
          & R4 \& G2 & 1     & 0.88  & 0.75  & 0.63 \\
    \hline
    \end{tabular}%
    \end{small}
    
    \caption{$pDEI$ Scores in Industry $S_6$. }
    \label{tab:s6_uni}
\end{table}


\section{Conclusion}

\begin{figure}[htbp]
    \centering
      \begin{subfigure}{0.4\columnwidth}
        \centering
        {\hspace*{-1.5cm}\includegraphics[width = 2\columnwidth, trim=6cm 3cm 0cm 2cm, clip]{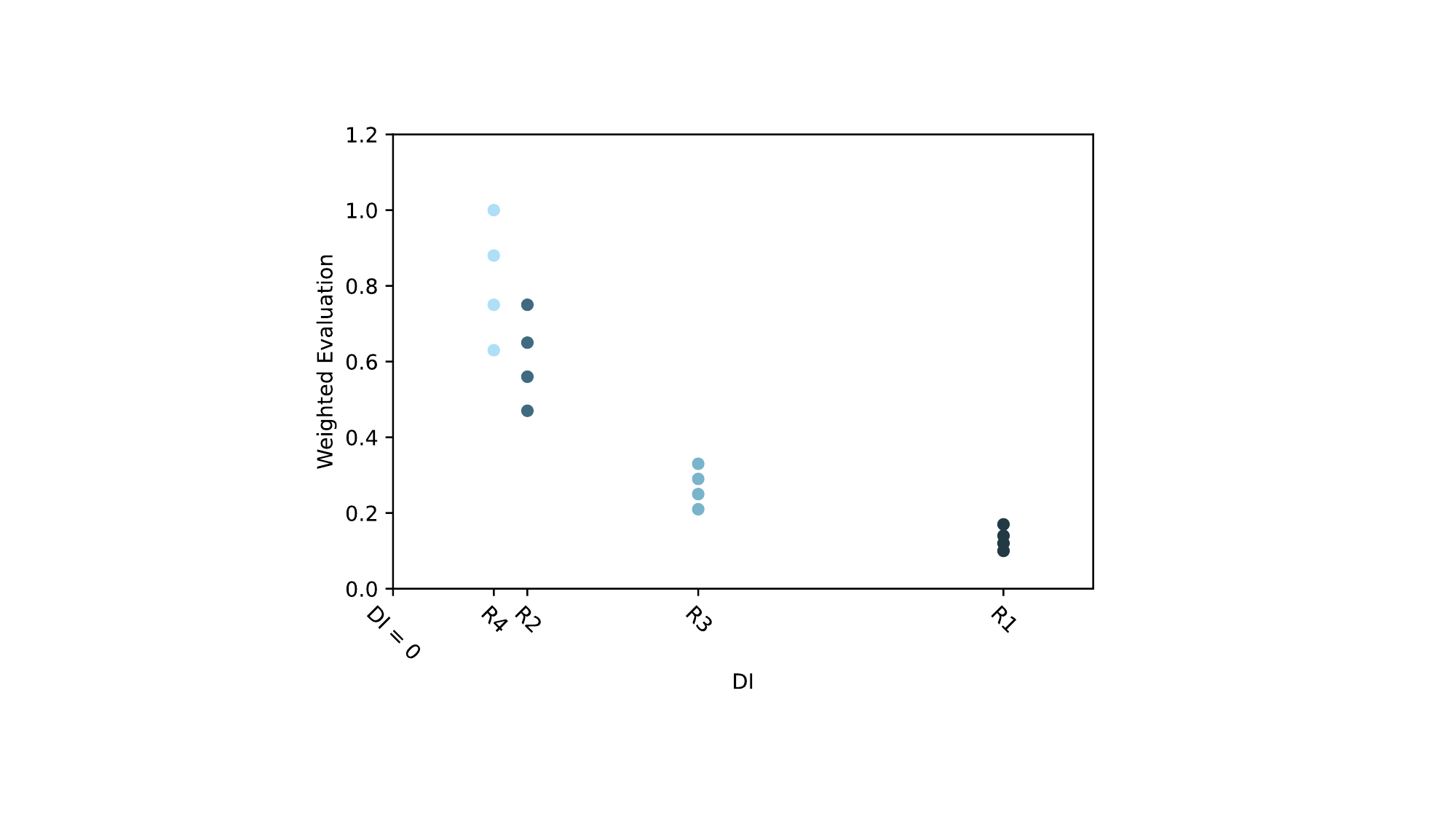}
        \caption{Industry $S_1$}}
        \label{fig:s-1}
      \end{subfigure}
      \begin{subfigure}{0.4\columnwidth}
        \centering
        \includegraphics[width = 2\columnwidth, trim=6cm 3cm 0cm 2cm, clip]{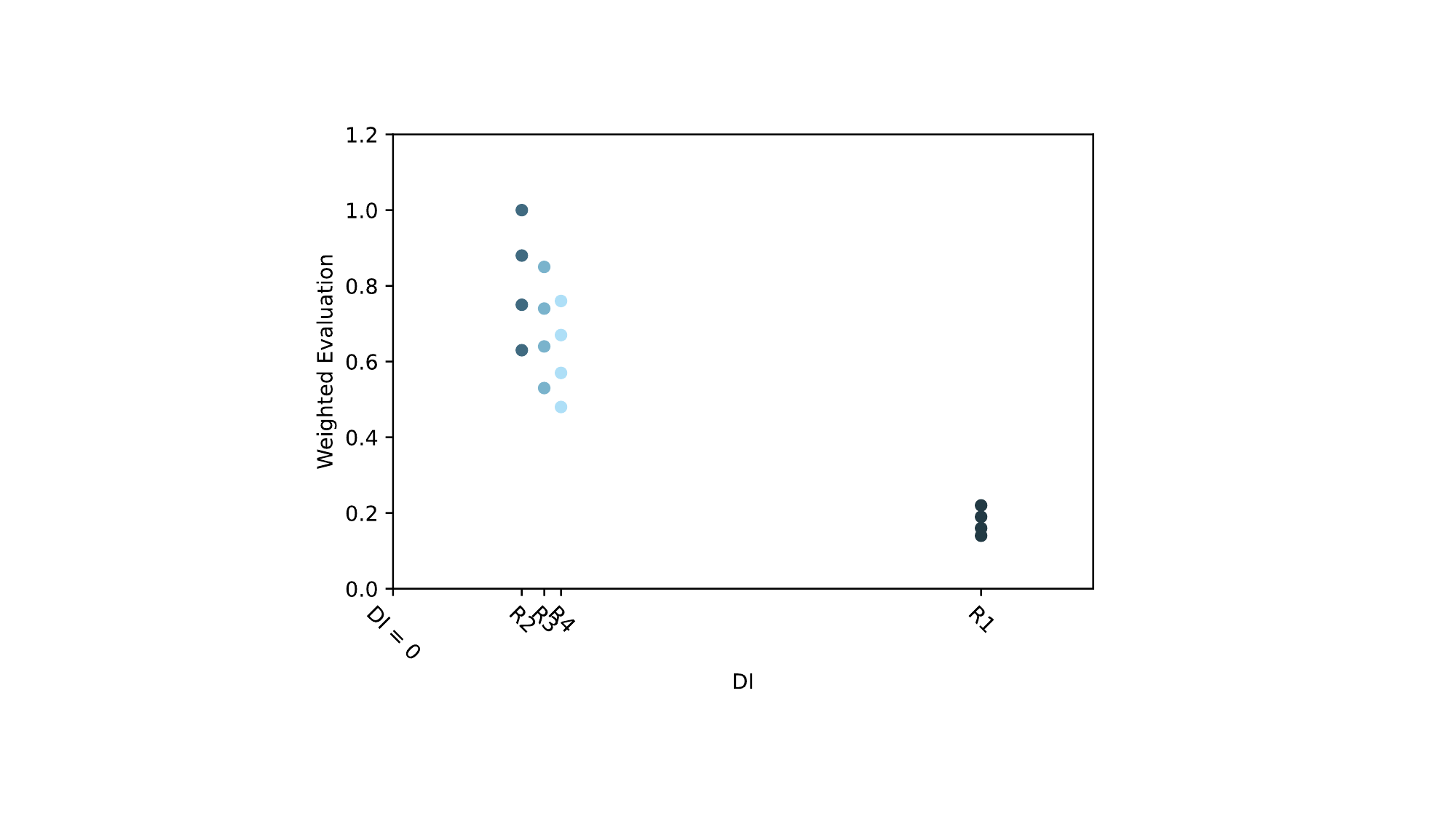}
        \caption{Industry $S_2$}
        \label{fig:s-2}
      \end{subfigure} 
      
     \begin{subfigure}{0.4\columnwidth}
        \centering
        {\hspace*{-1.5cm}\includegraphics[width = 2\columnwidth, trim=6cm 3cm 0cm 2cm, clip]{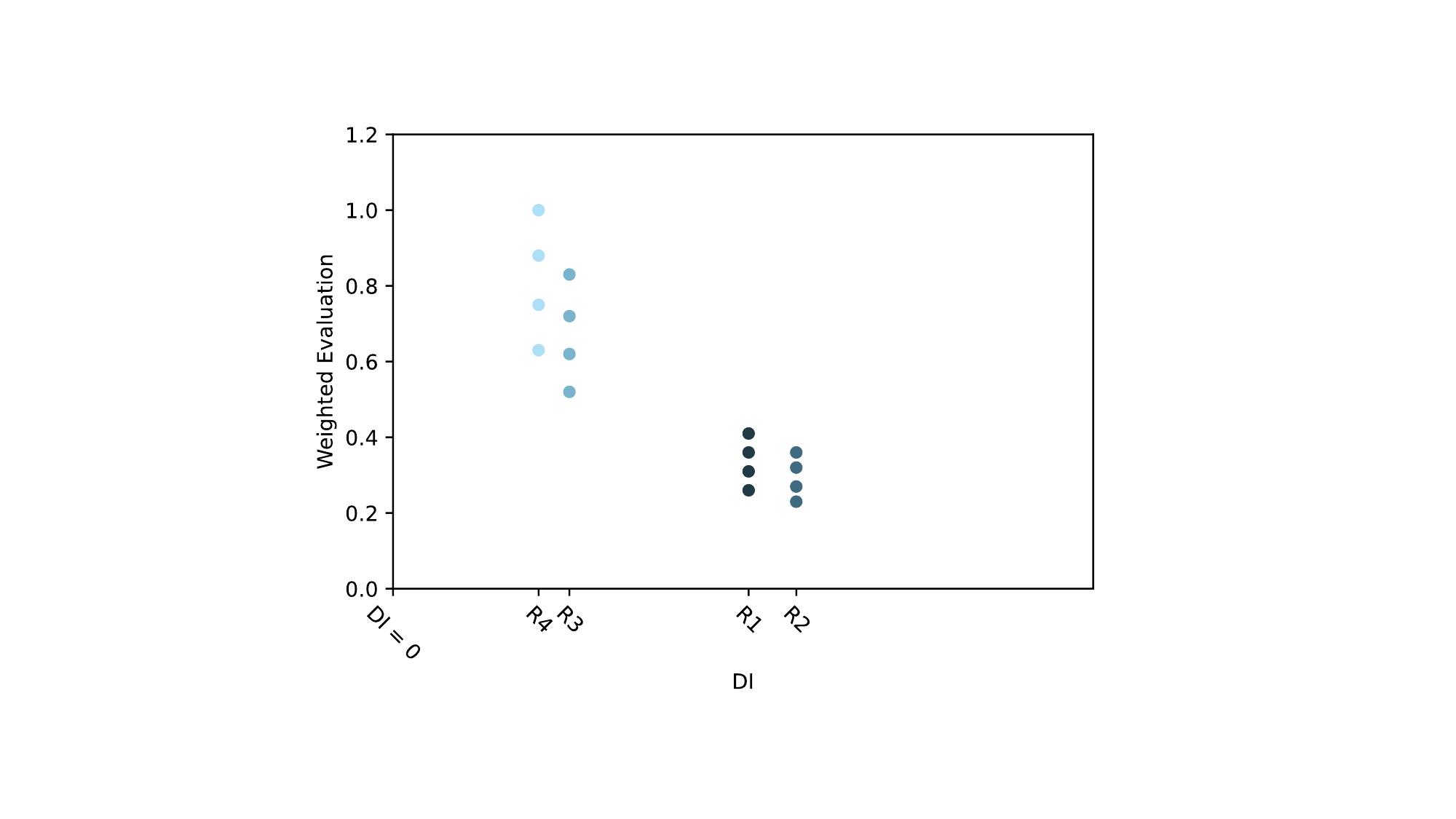}
        \caption{Industry $S_5$}}
        \label{fig:s-3}
      \end{subfigure}
      \begin{subfigure}{0.4\columnwidth}
        \centering
        \includegraphics[width = 2\columnwidth, trim=6cm 3cm 0cm 2cm, clip]{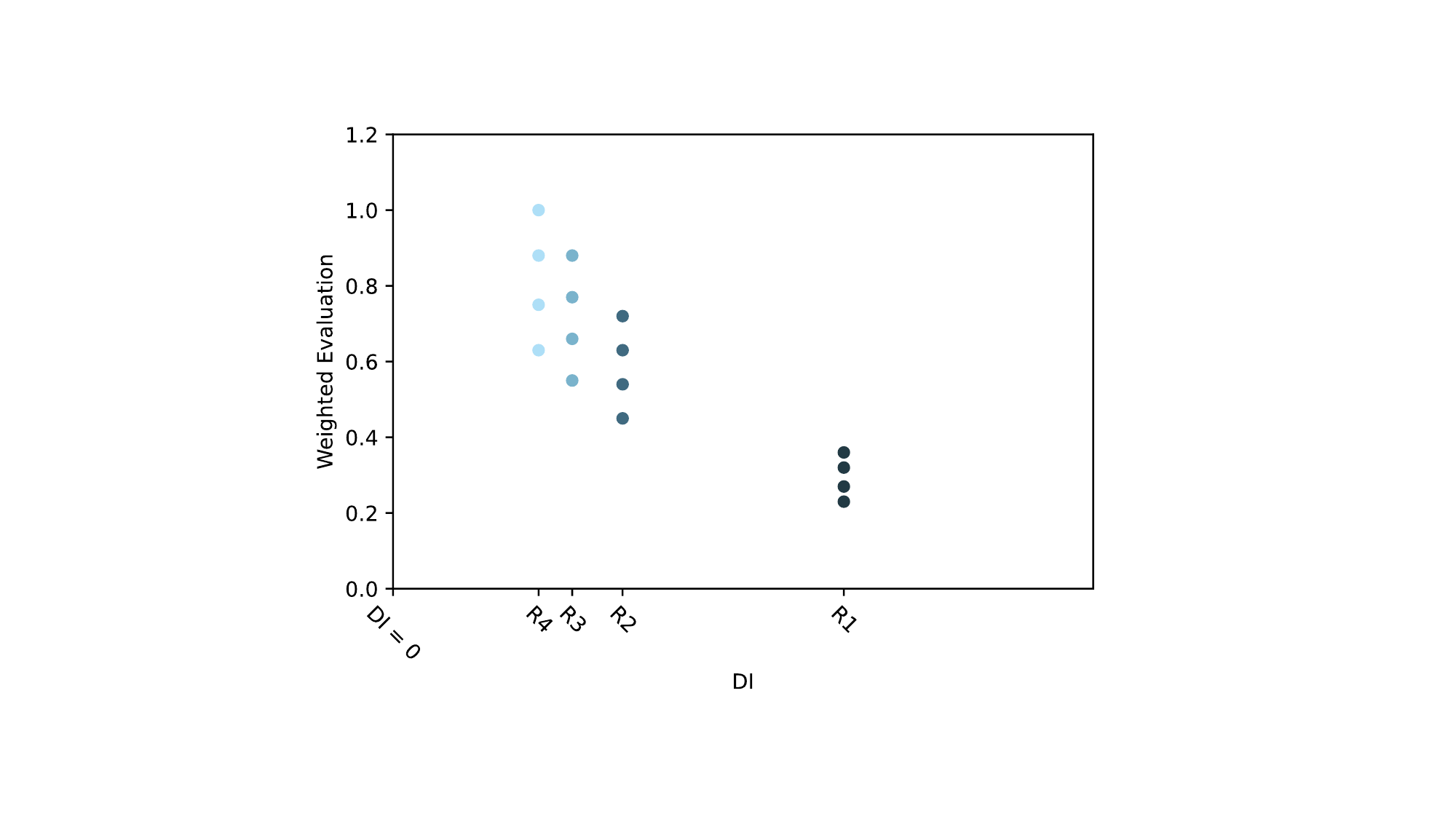}
        \caption{Industry $S_6$}
        \label{fig:s-4}
      \end{subfigure} 
    \caption{Adjusted Performance in Different Industry Sector adjusted by Race and Ethnicity.}
    \label{fig:con}
\end{figure}

\begin{figure}[htbp]
    \centering
      \begin{subfigure}{0.4\columnwidth}
        \centering
        {\hspace*{-1cm}\includegraphics[width = 2.3\columnwidth, trim=10cm 3cm 0cm 2cm, clip]{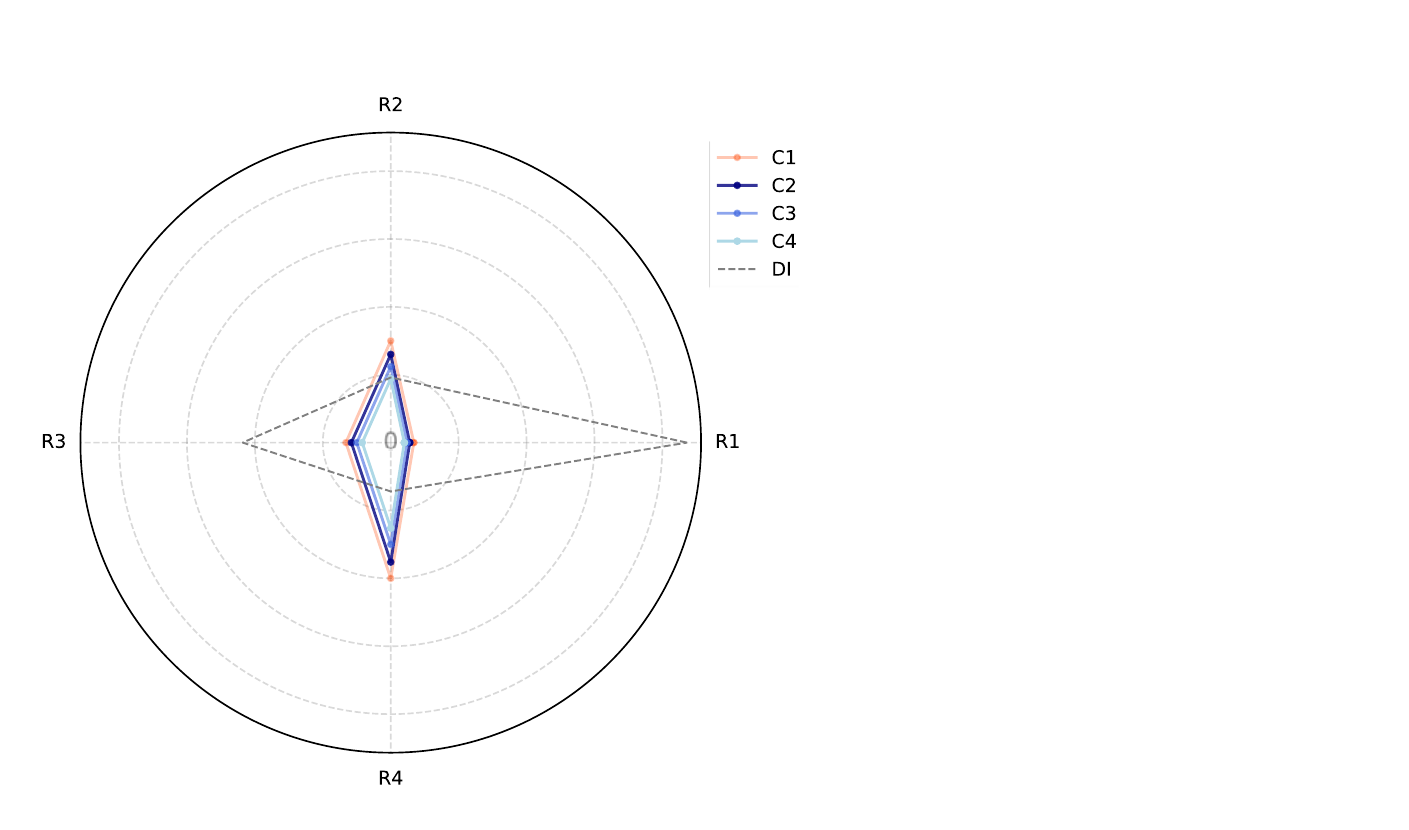}
        \caption{Industry $S_1$}}
        \label{fig:ss-1}
      \end{subfigure}
      \begin{subfigure}{0.4\columnwidth}
        \centering
        {\hspace*{-0cm}\includegraphics[width = 2.3\columnwidth, trim=10cm 3cm 0cm 2cm, clip]{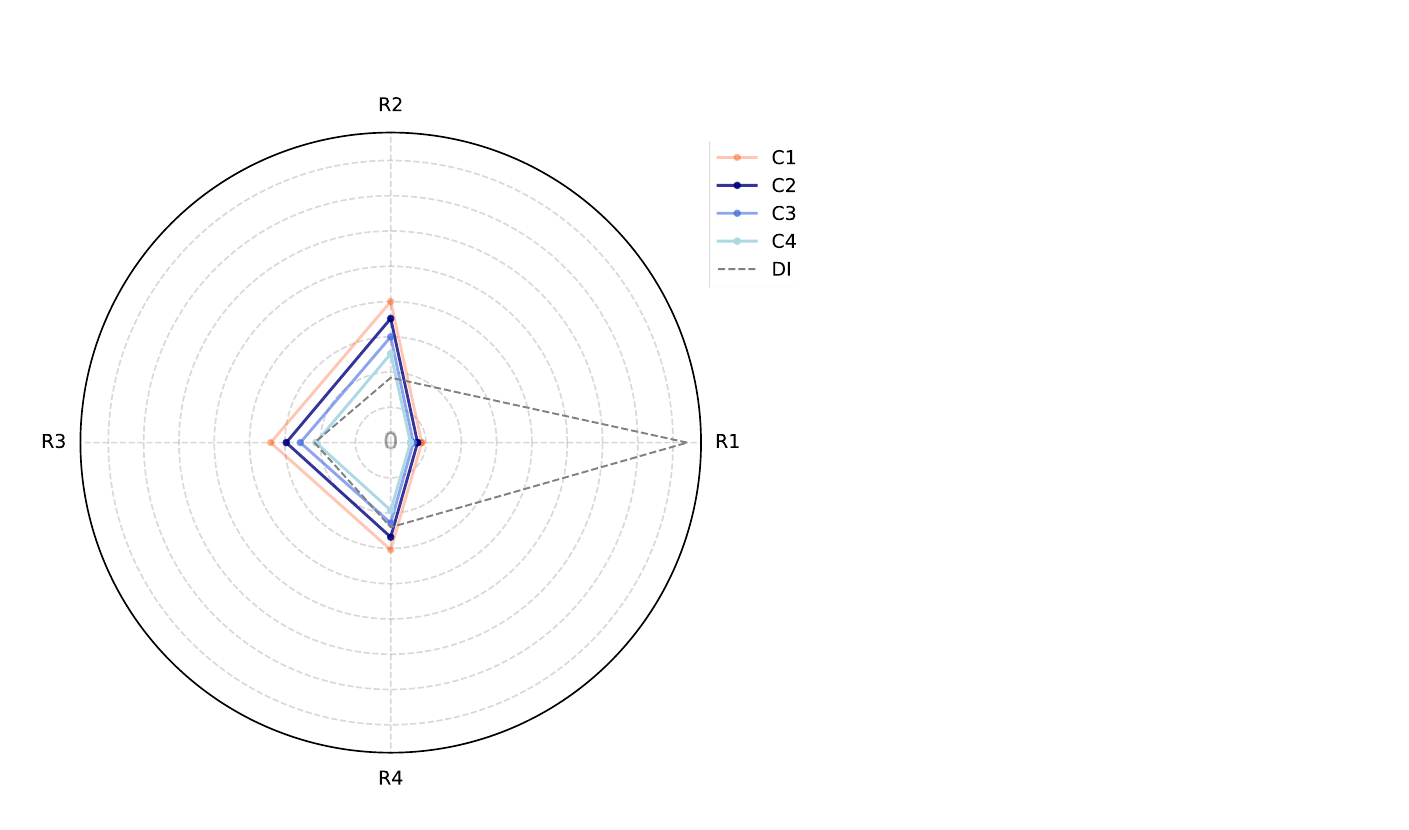}
        \caption{Industry $S_2$}}
        \label{fig:ss-2}
      \end{subfigure} 
      
     \begin{subfigure}{0.4\columnwidth}
        \centering
        {\hspace*{-1cm}\includegraphics[width = 2.3\columnwidth, trim=10cm 3cm 0cm 2cm, clip]{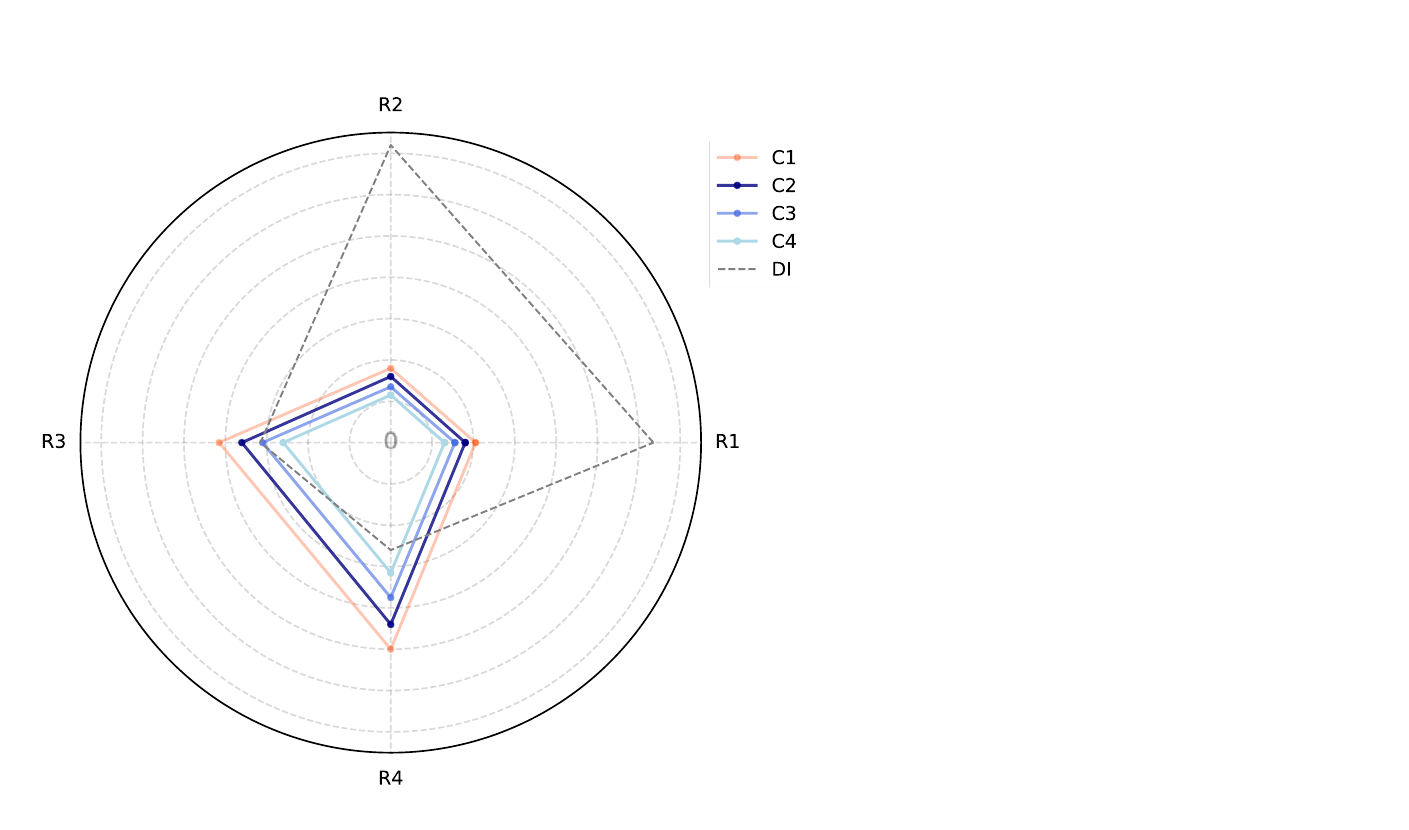}
        \caption{Industry $S_5$}}
        \label{fig:ss-3}
      \end{subfigure}
      \begin{subfigure}{0.4\columnwidth}
        \centering
        {\hspace*{-0cm}\includegraphics[width = 2.3\columnwidth, trim=10cm 3cm 0cm 2cm, clip]{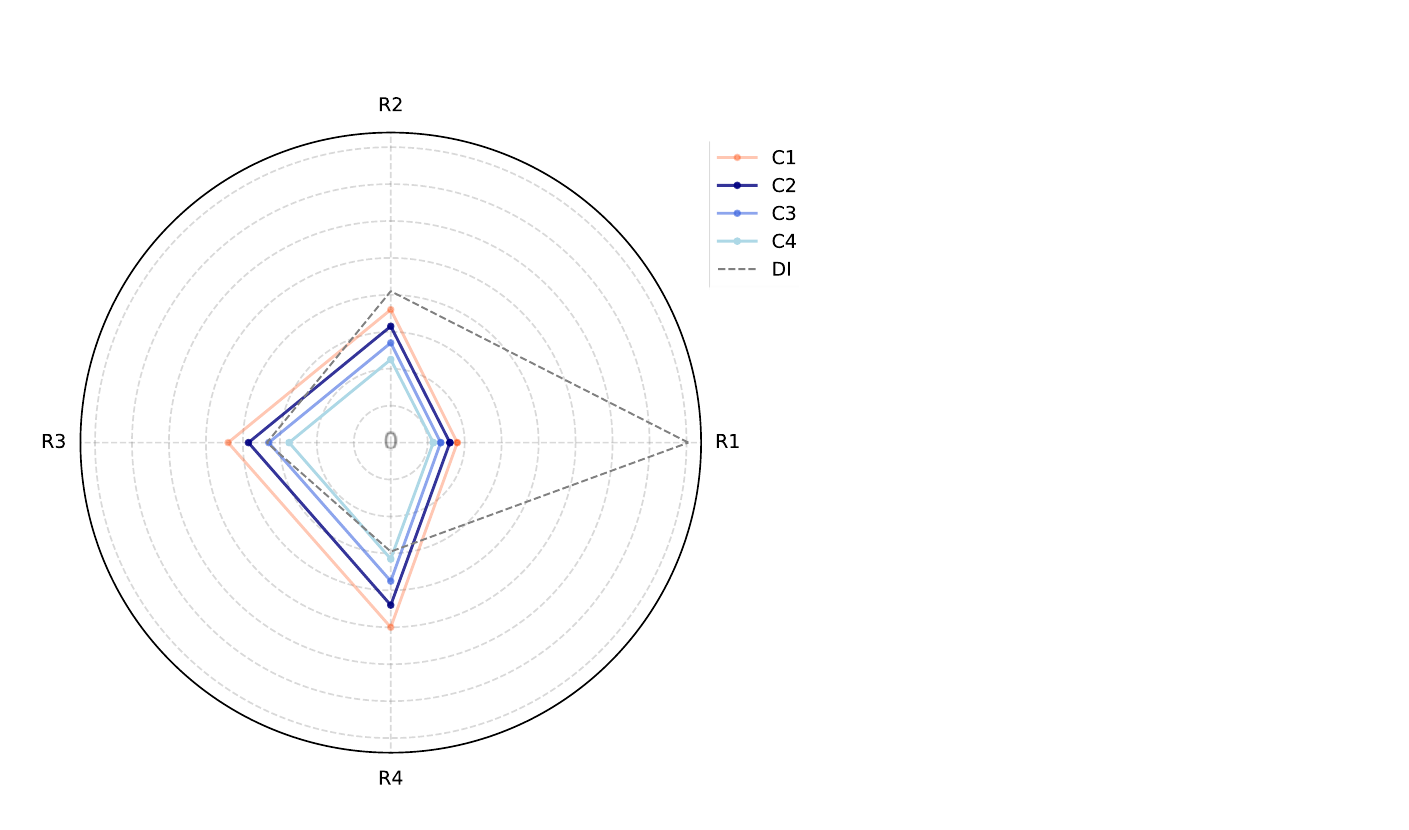}
        \caption{Industry $S_6$}}
        \label{fig:ss-4}
      \end{subfigure} 
    \caption{Adjusted Performance in Different Industry Sector adjusted by Race and Ethnicity.}
    \label{fig:conStar}
\end{figure}

Drawing from the numerical findings, it is evident that the DEA model incorporating the $pDEI$ metric offers a means to calibrate candidate performance in accordance with their socioeconomic disparity status. This adjustment can be primarily ascribed to the relative disparities observed within their DI metrics, as pictured in Figures \ref{fig:con} and \ref{fig:conStar}.

Figure \ref{fig:con} visually represents the final adjusted scores through a scatter plot, where the x-axis spans DI values ranging from the smallest (0, indicating the most underrepresented) to the maximum within those sectors, as derived from the data presented in Table \ref{tab:DI}. 

Similarly, Figure \ref{fig:conStar} consolidates the scatter plots into a polar coordinate system, with reference to the DI metrics represented by the dots connected by dotted lines. From those visualization, the model do underscore that groups with lower fairness scores should contemplate a more substantial degree of scaling adjustment -- even when cross-evaluating candidate performance -- in order to augment their DEI-adjusted efficiency scores. As depicted in Figure \ref{fig:conStar}, when a group exhibits a higher degree of overrepresentation relative to other groups (indicated by a sharp peak in the dotted line), a corresponding substantive degree of compensation should be allocated to underrepresented groups as a counterbalance to the overrepresented scores.

This observation could serve as a guiding principle for reweighting scores for each group within AI-driven decision tools aimed at constructing interpretable machine learning models. Furthermore, through a comparative analysis of the DEI-adjusted scores presented in Tables \ref{tab:s1_uni}, \ref{tab:s2_uni}, \ref{tab:s5_uni}, and \ref{tab:s6_uni}, it becomes evident that determined by the DI metric, one dimension of the input may dominate the others. For instance, in Table \ref{tab:s1_uni}, when simultaneously considering race-based DI and gender-based DI, as observed in Sector 1 where women are extremely underrepresented, all candidates within the ``Woman" group (R1\&G1, R2\&G1, R3\&G1, and R4\&G1) are evaluated with the same degree of efficiency, irrespective of whether they are overrepresented concerning their race or not. 

A plausible explanation for this phenomenon lies in the impact of input metric scales on model outcomes. While normalization is commonly employed to mitigate scale imbalances in machine learning and deep learning algorithms, in case of DEI-related questions, basic normalization techniques might be inadequate in capturing these disparities. However, without appropriate normalization, certain metrics may exert a dominant influence over others, rendering the trained model susceptible to issues such as memorizing ``model short-cut" and diminishing interpretability.

\section{Future Improvements}

As far as the authors are aware, the predominant body of existing research places a predominant emphasis on assessing fairness through the lens of equal treatment rather than delving into the intricacies of DEI considerations. Addressing how models can effectively compensate underrepresented groups poses a considerably complex challenge.

Through the utilization of the proposed $pDEI$ metric, we have observed intriguing patterns that can guide AI practitioners in reevaluating and interpreting their models in the context of fairness and disparity considerations. 

Nonetheless, it is important to acknowledge the presence of certain limitations that we intend to address more comprehensively in future research.  One such limitation, as highlighted in the conclusion, pertains to the redistribution of scores from overrepresented groups to underrepresented groups, with the extent of transfer increasing as the degree of overrepresentation grows. As illustrated in Table \ref{tab:s1_uni}, when solely considering the race-based DI, the top candidate from group $R1$ sees their score reduced to 0.17, rendering this candidate less favorable than the majority of candidates from other demographic groups. This phenomenon arises from the considerably high race-based DI value for $R1$ at 2.18, in contrast to the DI value of 0.48 for the most underrepresented group. This substantial disparity in DI values leads the model to substantially reduce the $pDEI$ scores for individuals within the $R1$ group. This observation sheds light on a potential research area, namely, the incorporation of constraints on utility transfers from overrepresented to underrepresented groups in the applied model.


\bibliography{aaai24}

\end{document}